\documentclass[12pt]{iopart}
\usepackage{color, soul}
\usepackage[pdftex]{graphicx}
\graphicspath{{./}}
\usepackage[outdir=./]{epstopdf}
\usepackage{url}

\begin{document}

\title{High Coherence Plane Breaking Packaging for Superconducting Qubits}

\author{Nicholas T. Bronn$^1$, Vivekananda P. Adiga$^1$, Salvatore B. Olivadese$^1$, Xian Wu$^2$, Jerry M. Chow$^1$, David P. Pappas$^2$}

\address{$^1$IBM T.J. Watson Research Center, Yorktown Heights, NY 10598, USA}
\address{$^2$National Institute of Standards and Technology, Boulder, CO 80305, USA}
\ead{ntbronn@us.ibm.com}

\vspace{10pt}

\begin{abstract}
We demonstrate a pogo pin package for a superconducting quantum processor specifically designed with a nontrivial layout topology ({\it e.g.}, a center qubit that cannot be accessed from the sides of the chip). Two experiments on two nominally identical superconducting quantum processors in pogo packages, which use commercially available parts and require modest machining tolerances, are performed at low temperature (10~mK) in a dilution refrigerator and both found to behave comparably to processors in standard planar packages with wirebonds where control and readout signals come in from the edges. Single- and two-qubit gate errors are also characterized via randomized benchmarking, exhibiting similar error rates as in standard packages, opening the possibility of integrating pogo pin packaging with extensible qubit architectures.
\end{abstract}

%
\vspace{2pc}

\section*{Introduction}

Superconducting qubits have become one of the leading candidates for building near-term quantum computing processors, enabling demonstrations of various aspects of quantum error correction~\cite{Corcoles2015,Kelly2015,Riste2015,Ofek2016}, short-depth variational approximate quantum algorithms~\cite{OMalley2016,Kandala2017}, and even cloud access~\cite{qx}. Yet, in all such experiments, control and readout signals are routed to the qubit processor along the edge of the chip from traces on a printed circuit board (PCB) package. This packaging technique is physically limiting to the arrangement of qubits on the processor. For example, with topologies as required for logically encoding within a surface code~\cite{Fowler2012,Gambetta2017,Versluis2016}, it becomes necessary to address qubits which might be surrounded on all sides by other qubits or circuit elements such as microwave quantum buses or readout resonators.

One such method of overcoming this topological restriction is to route the signals out of the plane of the qubit device~\cite{Gambetta2017}, for example using a three-dimensional quantum socket~\cite{Bejanin2016}, through-silicon vias~\cite{Versluis2016}, flip-chip multi-layer stack~\cite{Rosenberg2016}, micromachined cavities~\cite{Brecht2017}, or multi-layered printed circuit boards~\cite{Liu2017}. In this article, we present a technique for breaking the plane using physical pogo pins inside dielectric plugs which form a 50~$\Omega$ coaxial connection when mounted in an interposer between a circuit board and the qubit chip. An impedance of 50~$\Omega$ is chosen for compatibility with radio frequency (RF) instruments without using impedance transformers. This also offers the advantage of using commercially available parts and modest machining requirements. The careful assembly of such packages is critical in obtaining compatibility with quantum measurements. 

We show that for a nontrivial lattice of seven superconducting transmon qubits that the pogo package performs as well as a standard PCB with regards to qubit coherence time and control fidelities. An additional advantage to this approach is that it can be combined with other integration techniques such as on-chip wirebonds or crossovers for spurious microwave mode supression. It is also compatible with rapid turn-around and test of devices and potential post-measurement device modification. 

\section*{Methods} 

\begin{figure}[!h]
	\centering
	\includegraphics[width=5.5in]{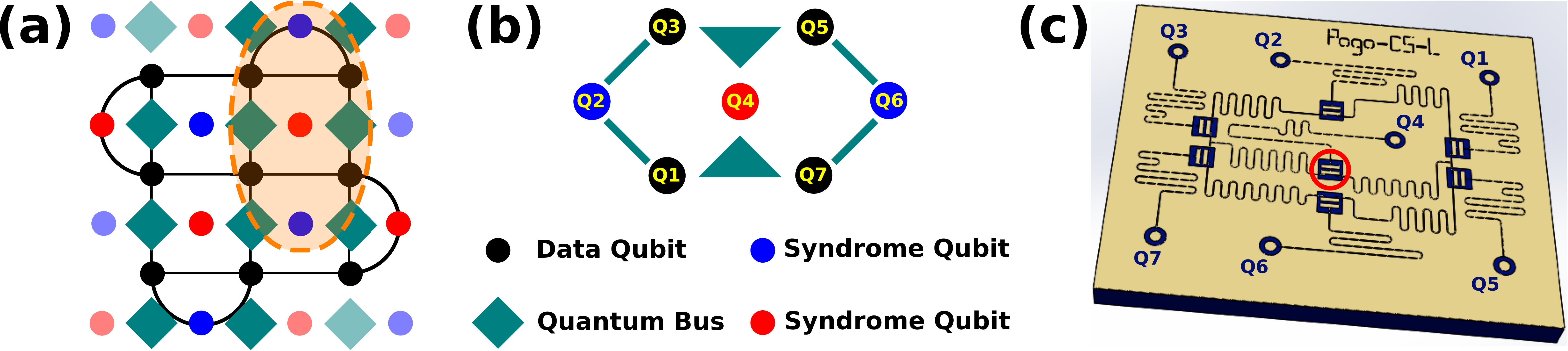}
	\caption{\label{fig:surface-code} \textbf{Surface Code of Quantum Error Correction.} (a) Layout of the rotated surface code with quantum buses (diamonds) mediating interactions between data qubits and syndrome qubits. (b) Layout of the nontrivial 7-qubit circuit topology, where Q4 is the qubit in the center. (c) Computer-aided design schematic of the quantum processor, with a red circle indicating Q4.}
\end{figure}

We choose a non-trivial arrangement of seven qubits which requires our pogo technique to break the plane. The qubit layout is shown in Fig.~\ref{fig:surface-code}a and happens to be a particular cut-out of the rotated surface code. A different seven-qubit cut-out of the rotated surface code has been studied before~\cite{Takita2016}, but in this particular arrangement, there are six quantum buses (two triangles connecting 3 qubits each and four lines connecting 2 qubits each, as shown in Fig.~\ref{fig:surface-code}b) to mediate the interaction between qubits, surrounding the center qubit (Q4) such that it cannot be accessed from the edge of the chip with a wirebond. We design this arrangement with the frequencies of the readout resonators, bus resonators and qubits targeting $6.7-7.0$~GHz, $5.9-6.4$~GHz, and 5~GHz, respectively, and with coupling strengths $J/2\pi \approx 3$~MHz to enable nearest-neighbor two-qubit gates.

\begin{figure}[!h]
	\centering
	\includegraphics[width=5.5in]{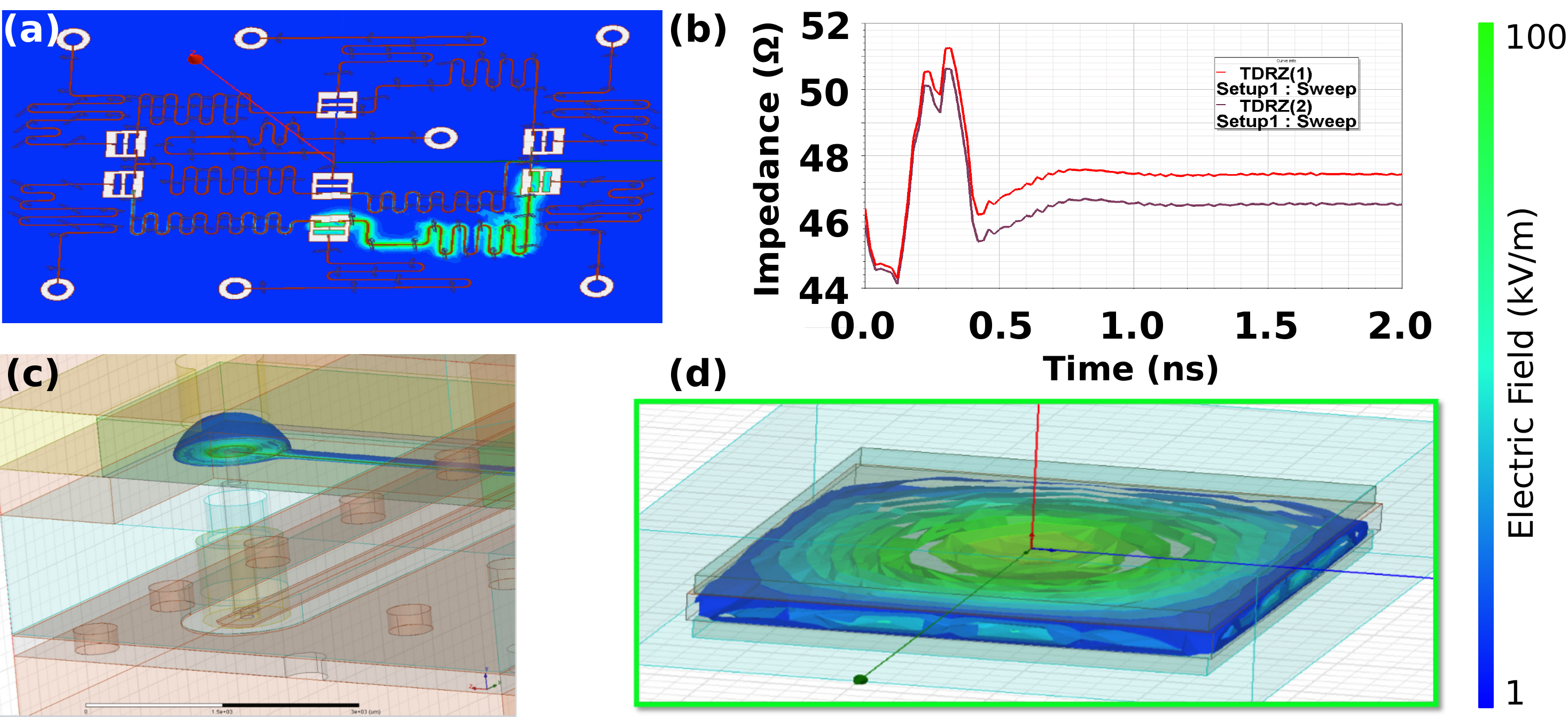}
	\caption{\label{fig:hfss-sims} \textbf{High-frequency Microwave Simulations.} (a) An example of a single resonance at a given resonator frequency. (b) Simulated TDR shows small deviations from 50~$\Omega$ impedance at the pogo pin/PCB via transition between 0 and 0.5~ns. TDR from port 1 (the pogo pin/teflon/interposer coaxial connection) to port 2 is in red, while the TDR from port 2 to port 1 (in purple, co-planar waveguide on the quantum processor) is reflected in time for comparison with the red curve. (c) Model for the pogo pin-launch interface simulated in (b). (d) The fundamental package mode of the quantum processor die and package with a recess below the die. }
\end{figure}

Microwave simulations are important to help design the quantum processor, pogo pin-launch interface, and pogo package. These simulations indicate that the resonator frequencies are sufficiently spaced apart so only a single one is excited at each frequency (Fig.~\ref{fig:hfss-sims}a).  The pogo pin-launch interface is also simulated (Fig.~\ref{fig:hfss-sims}c) to determine that a nearly 50~$\Omega$ transition, as verified by time domain reflectometry (TDR) shown in Fig.~\ref{fig:hfss-sims}b, can be made if a circular pad of diameter 300~$\mu$m with an antipad of diameter 550~$\mu$m is used to capture the signal from the pogo pin. The size of the quantum processor die is $9.5 \times 11.5$~mm$^2$, larger than our original seven-qubit device~\cite{Takita2016}. Qubit or resonator coupling to the fundamental package mode is a limiting feature for larger and larger devices until the development of cryogenically-compatible via technology~\cite{Wenner2011, Gambetta2017}. Initial simulations indicate that the frequency of the fundamental package mode of the die (with a 0.5~mm vacuum gap above) is at 9.0~GHz. However, if a recess is made below the die, the fundamental package mode is pushed up to 13.4~GHz (Fig.~\ref{fig:hfss-sims}d), far higher than the frequencies of the qubits and resonators. Following these design considerations informed by microwave simulation, the quantum processor is fabricated using standard techniques~\cite{Corcoles2015}.

\begin{figure}[!h]
	\centering
	\includegraphics[width=5.5in]{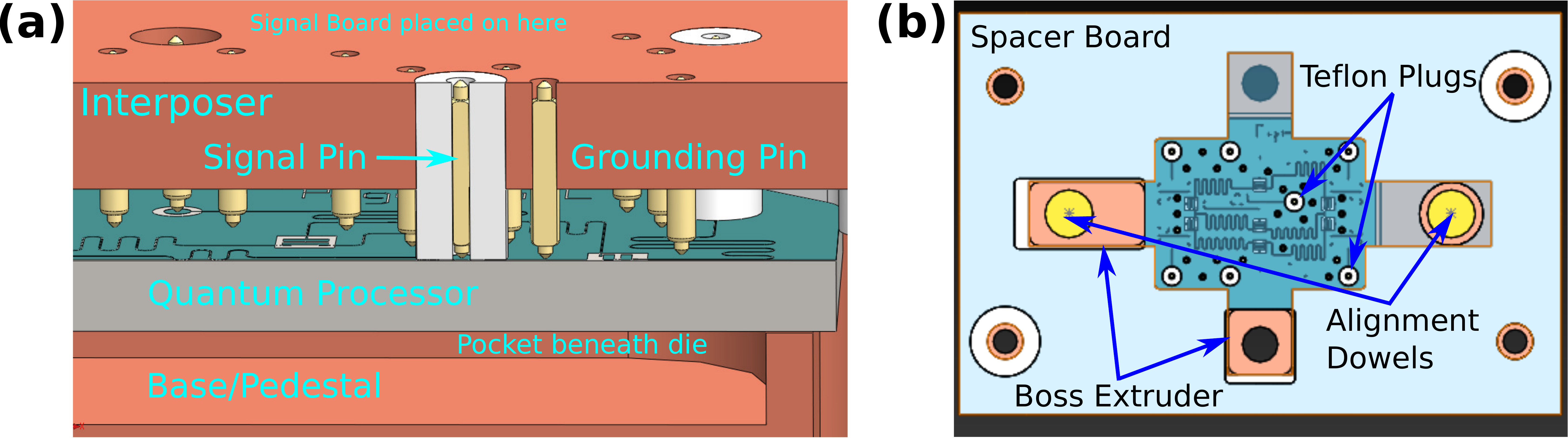}
	\caption{\label{fig:pogo-xsection} \textbf{Computer-aided Design Views.} (a) Cross section of the pogo package shows pogo pins, which rest in teflon plugs held in place by the interposer, that route signals between a PCB (not shown) and capture pads on the quantum processor. Other pogo pins may be used for ground connections. The quantum processor rests on the base/pedestal and is held down by the spacer board (not shown). (b) View from above of the pogo packaging shows the quantum processor held down by the spacer board, boss and dowels used for alignment, and teflon plugs holding pogo pins (interposer not shown).}
\end{figure}

\begin{figure}[!h]
	\centering
	\includegraphics[width=5.5in]{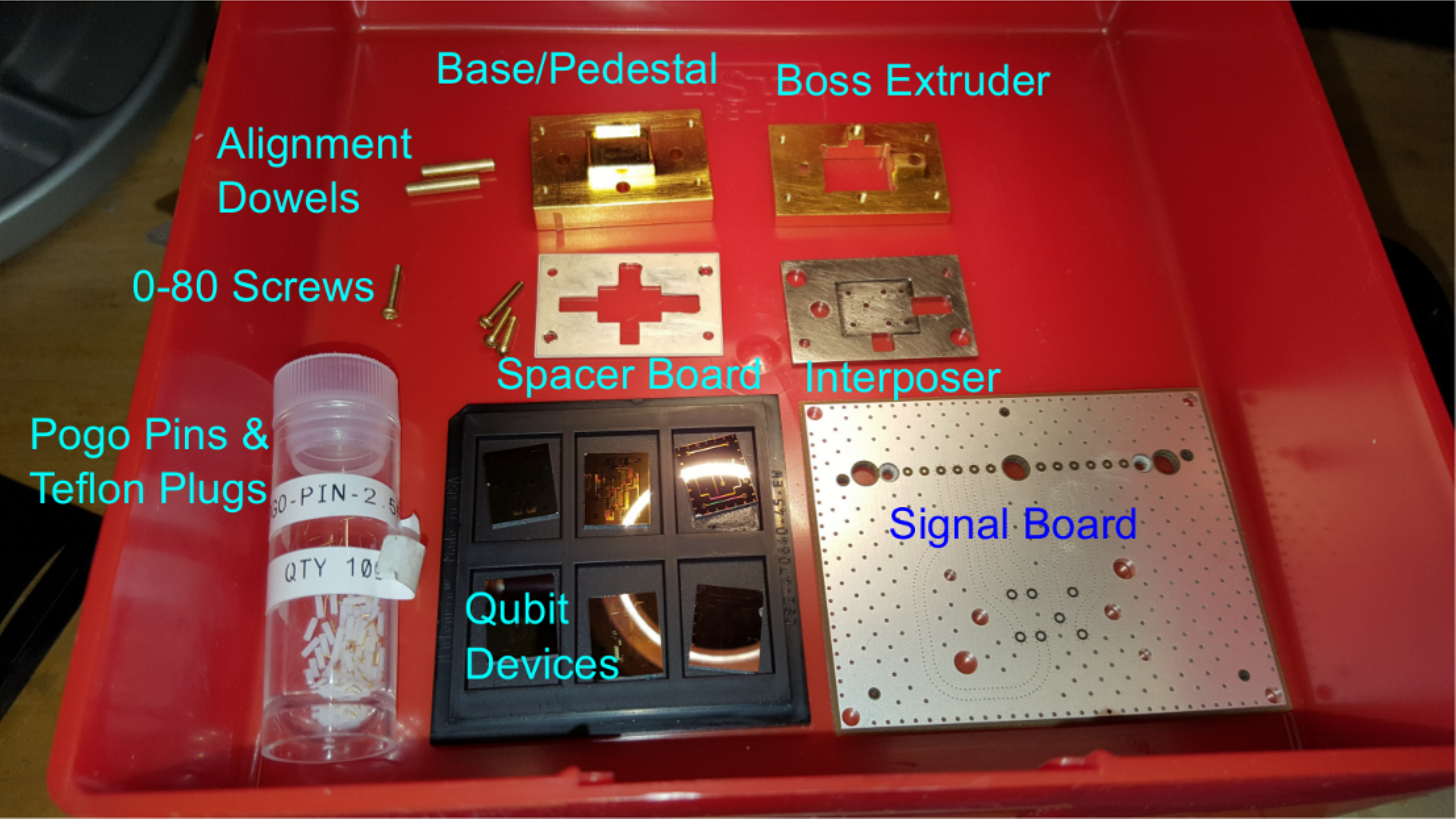}
	\caption{\label{fig:pogo-components} \textbf{Pogo Package Components.} The high-throughput connector and backing plate are not shown. The base/pedestal consists of gold-plated copper, the spacer board is a 20 mil silver-plated duroid printed circuit board, and the signal board is a 4-layer silver-plated FR-4 board. The boss extruder is made of aluminum and interposer of gold-plated copper, while the composition of the shown parts is different. Grounding pins, the holes for which surround the large signal holes in the interposer, were not used in the experiments.}
\end{figure}

The package is designed for pogo pins to take signals from a printed circuit board to and from capture pads designed on the quantum processor while maintaining an impedance of 50~$\Omega$. Other pogo pins may be used as ground connections as well. A computer-aided design cross section of this package is displayed in Fig.~\ref{fig:pogo-xsection}. The components of the pogo package are shown in Fig.~\ref{fig:pogo-components}. It consists of a gold-plated copper base that is thermally anchored to the coldfinger of a dilution refrigerator while the quantum processor sits on a pedestal rising from the base, together called the base/pedestal. As informed by microwave simulations discussed previously, the pedestal has a pocket milled out underneath the quantum processor in order to push undesirable package modes to higher frequencies. The aluminum boss extruder aligns to the base/pedestal via brass alignment dowels and is used to align the other components.  A 20~mil thick duroid PCB coated in copper and plated in silver makes contact with the ground planes of the quantum processor and holds it into place with the base/pedestal. The gold-plated copper interposer holds the pogo pins and teflon dielectric plugs in place so that they align to and mate with the signal launches on the qubit device and signal board. Cylindrical dielectric plugs holds each pogo pin in place, and the size of the teflon plugs carrying the signal ensures a $50~\Omega$ impedance, so that each pogo pin/teflon plug/interposer acts as a coaxial transmission line, as in Fig.~\ref{fig:hfss-sims}c-d. The pogo pins are made from a gold-plated hardened beryllium copper alloy with a maximum diameter of 0.31~mm, full travel length of 0.6~mm and recommended (compressed) travel length of 0.45~mm~\cite{pogo-specs}. The signal board is a four-layer silver-plated PCB that takes the signals from pogo pins incident on capture pads to internal buried stripline transmission lines and back to the surface where it mates with a high-throughput commercial connector. A copper backing plate with copper braids that attach to the coldfinger sits atop the signal board to provide thermalization for the board and connector, and the connector also screws into this plate.

\begin{figure}[!h]
	\centering
	\includegraphics[width=5.5in]{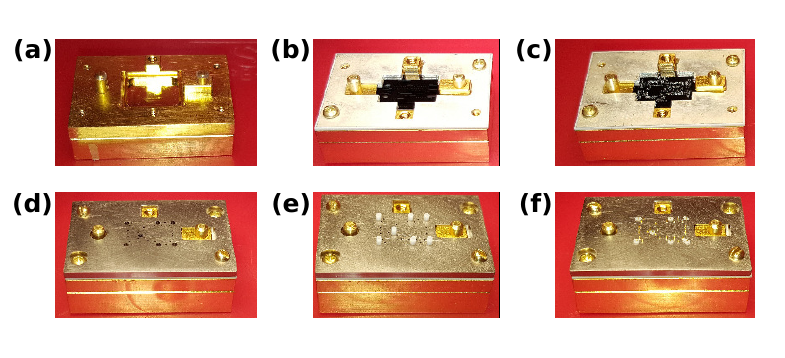}
	\caption{\label{fig:pogo-assembly} \textbf{Pogo Package Assembly.} (a) The base/pedestal and boss extruder attached with the alignment dowels. (b) The spacer board holds the quantum processor in place. (c) A wirebonded qubit device. (d) The interposer is attached and aligned by the boss extruder (one can observe the capture pads on the quantum processor through the large holes). (e) Dielectric plugs pushed into holes and resting on the qubit device. (f) Trimmed dielectric plugs populated with pogo pins.}
\end{figure}

The base/pedestal and boss extruder are aligned with two brass dowels, as in Fig.~\ref{fig:pogo-assembly}a. Future alignment steps are done with respect to either the dowels or the boss extruder. The spacer board is then secured against the quantum processor while it is pushed against the boss extruder (Fig.~\ref{fig:pogo-assembly}b). The screws that hold the spacer board to the boss extruder are placed in the diagonal where the larger holes in the interposer offer the screw heads clearance. The ground planes of the quantum processor are wirebonded together to prevent spurious slotline modes from coupling to the qubits and resonators (Fig.~\ref{fig:pogo-assembly}c)~\cite{Koch2007}. The interposer is then secured to the package with screws along the opposite diagonal while pushing it against the boss extruder. Dielectric plugs are then inserted into the seven large holes and pushed against the capture pads on the quantum processor. The plugs are then cut to the level of the interposer with a precision knife and populated with pogo pins. The screws along the original diagonal are removed and the pogo package is aligned with a signal board and a backing plate by the brass dowels. The removed screws are then used to secure the pogo package through holes in the backing plate. The alignment dowels are removed and long screws inserted in their place, holding the combined package together more securely with hex nuts. The backing plate is also thermally anchored to the coldfinger of the dilution refrigerator with copper braid.

\section*{Results}

Experiments performed with quantum processors are considered using nominally identical packaging, Pogo1  and Pogo2. The experiments were performed in the same dilution refrigerator and experimental setup, with the exception of input attenuation chain. The following data are for pogo packages with aluminum boss extruders and gold-plated copper interposers. Grounding pins were not present in either experiment. The qubits were measured in reflection by one of three input lines, each possessing a cryogenic ferrite switch so that two-qubit gates could potentially be performed between each pair of coupled qubits. Tables~\ref{table:pogo1coho}-\ref{table:pogo2coho} show representative qubit parameters. Two-qubit gates, composed of cross resonance pulses~\cite{Ashhab2006,Ashhab2007,Ashhab2008,Rigetti2010,Chow2011}, were tuned up between a few pairs of qubits~\cite{Sheldon2016a}. 

\begin{table}[h!]
	\centering
	\begin{tabular}{||l| c c c c c c ||} 
		\hline
		  & $f_{R}$ (GHz) & $f_q$ (GHz) & $Q$ & $T_2^{\rm echo}$ ($\mu$s) & EPC (RB) & EPC (Limit)  \\ [0.5ex] 
		\hline\hline
		Pogo1 Q1 & 6.8753 & 4.3012 & 2270000 & 41.0 & 0.00512(2.9e-4) & 0.00126 \\
		Pogo1 Q2 & 6.7297 & 4.7673 & 2060000 & 41.1 & 0.00370(1.6e-4) & 0.00175 \\
		Pogo1 Q3 & 6.9033 & 4.9912 & 1490000 & 10.6 & 0.00448(9.9e-5) & 0.00435 \\ 
		Pogo1 Q4 & 6.9728 & 4.7404 & 2180000 & 95.3 & -               & 0.00574 \\ 
		Pogo1 Q5 & 6.8332 & 5.0391 & 1540000 & 79.0 & 0.00302(8.6e-5) & 0.00096 \\ 
		Pogo1 Q6 & 6.7574 & 4.8277 & 2200000 & 23.3 & 0.00750(2.4e-4) & 0.00413 \\ 
		Pogo1 Q7 & 6.9370 & 4.9889 & 1670000 & 86.8 & 0.00478(1.5e-4) & 0.00087 \\ [1ex] 
		\hline
	\end{tabular}
	\caption{\label{table:pogo1coho} \textbf{Pogo1 Experiment Parameters.} The frequency of each qubit and its readout resonator is given. Qubit quality factor $Q$ and coherence time $T_2^{\rm echo}$ are similar to those measured with standard packaging. EPC is determined by RB and can be compared to the coherence-limited EPC. EPC for Q4, the qubit in the center of the chip, is not provided because the qubit decoheres before a reasonable number of gates for RB can be performed due to the long gate length for Q4. }
\end{table}

\begin{table}[h!]
	\centering
	\begin{tabular}{||l| c c c c c c ||} 
		\hline
		  & $f_{R}$ (GHz) & $f_q$ (GHz) & $Q$ & $T_2^{\rm echo}$ ($\mu$s) & EPC (RB) & EPC (Limit) \\ [0.5ex] 
		\hline\hline
		Pogo2 Q1 & 6.8703 & 4.9961 & 1520000 & 31.9 & -                & 0.00069 \\
	    Pogo2 Q2 & 6.7223 & 5.1413 & 1730000 & 23.9 & -                & 0.00085 \\
		Pogo2 Q3 & 6.8976 & 4.7921 & 1470000 & 75.4 & -                & 0.00039 \\ 
		Pogo2 Q4 & 6.9656 & 4.8721 & 1220000 & 48.3 & 0.00932(7.5e-4)  & 0.00057 \\ 
		Pogo2 Q5 & 6.8247 & 4.9491 & 1000000 & 50.2 & -                & 0.00052 \\ 
		Pogo2 Q6 & 6.7499 & 4.8731 & 1310000 & 22.0 & -                & 0.00052 \\ 
		Pogo2 Q7 & 6.9264 & 4.9314 & 1450000 & 70.0 & 0.00236(9.0e-5)  & 0.00042 \\ [1ex] 
		\hline
	\end{tabular}
	\caption{\label{table:pogo2coho} \textbf{Pogo2 Experiment Parameters.} The same parameters shown in Table~\ref{table:pogo1coho} for the Pogo2 experiment. Simultaneous RB with Q4 and Q7 yields an EPC of 0.0242(0.0104).}
\end{table}

\section*{Discussion}

The Pogo1 experiment demonstrates quality factors $Q=\omega_q T_1$, where $\omega_q =2\pi f_q$ is the (angular) qubit frequency and $T_1$ is the measured qubit lifetime, and coherence times $T_2^{\rm echo}$ in line with qubits measured in standard packaging~\cite{qx,Takita2016,Gambetta2017}, and are exhibited in Table~\ref{table:pogo1coho}. Qubit $Q$'s are measured from 1.5-2.3 million while coherence times  range from $10 - 95$~$\mu$s. These variations in coherence are larger than expected from the qubit circuit itself. Further optimization of the package geometry is expected to reduce these variations. It is suspected that the boss extruder does not thermalize correctly due to the low thermal conductivity of aluminum below its superconducting critical temperature. Another possibility is the oxide of the aluminum makes an imperfect electrical seal as opposed to gold. The variation in $T_2^{\rm echo}$ values could also be indicative of differences in the connection between pogo pin and capture pad, leading to effective variance in line attenuation and thereby qubit dephasing. Further experiments will discern the effects of material on qubit lifetime and coherence by substituting parts of the pogo package. However, single-qubit gate lengths are quite long in the Pogo1 experiment (hundreds of nanoseconds), especially for the center qubit, Q4 (one microsecond), due to the experimental setup. Hence, for the Pogo2 experiment, the cryogenic attenuation profile is modified to allow faster gates. A nominally similar quantum processor and pogo package is used for this experiment, and the results can be found in Table~\ref{table:pogo2coho}. Qubit quality factors for the Pogo2 experiment are not quite as high as for Pogo1, and the coherence times are still in a rather large range, from 22-75~$\mu$s. However, the single-qubit gate lengths are much shorter, around 50~ns as is typical for standard packaging~\cite{Takita2016}. 

Error per Clifford (EPC) gate are determined by randomized benchmarking (RB) in both the Pogo1 and Pogo2 experiments~\cite{Magesan2012}. EPCs are in the range of 0.003-0.008 for the Pogo1 experiment and 0.002-0.01 for Pogo2, and specified in Table~\ref{table:pogo1coho} and Table~\ref{table:pogo2coho}, respectively. This compares to a coherence-limited EPC range of 0.001-0.006 for Pogo1 and 0.0004-0.0009 for Pogo2. Simultaneous single-qubit RB increases these EPCs to a range of 0.004-0.009 for Pogo1 and 0.0242(0.0104) for Pogo2, indicating that crosstalk limits the ability to perform simultaneous gates. The distinction between classical and quantum crosstalk that arises from qubit bus coupling and packaging is a topic of current research interest, and will be explored in further work. After changing the input line attenuation so that faster gates could be performed, a 400~ns two-qubit $ZX_{\pi/2}$ gate with an EPC of 0.0649(0.0014) is tuned up and performed between Q1 and Q2 of the Pogo2 experiment~\cite{Sheldon2016a}. 

\section*{Conclusion}

The ability to maintain high coherence qubits in pogo packages is demonstrated. Single- and two-qubit gates are also characterized. Further experiments are planned on similar iterations of the pogo package, for example, by adding grounding pins in subsequent packages and substituting parts of the pogo package with different materials. These experiments confirm that the pogo package represents a viable path forward for extensible qubit integration towards a logical qubit.

\section*{Acknowledgements}

The authors acknowledge experimental contributions from M. Brink, O. Jinka, S. Rosenblatt, M.O. Sandberg, S. Sheldon, and M. Takita. This work was supported by IARPA under contract W911NF-16-1-0114-FE. All statements of fact, opinion, or conclusions contained herein are those of the authors and should not be construed as representing the official views or policies of the U.S. Government. This work is property of the U.S. Government and not subject to copyright. Any mention of commercial products is for information only; it does not imply recommendation or endorsement by NIST. We acknowledge support of the NIST Quantum Information Initiative.

\section*{References}

\bibliographystyle{iopart-num}
\bibliography{paper-pogos7}

\end{document}